\documentclass[aps,prl,twocolumn,showpacs,preprintnumbers,amsmath,amssymb]{revtex4}
\usepackage{graphicx}
\usepackage{hyperref}
\usepackage{natbib}
\usepackage{dcolumn}
\usepackage{bm}

\begin{document}

\title{Magnetism and local distortions near carbon impurity in $\gamma$-iron}

\author{D. W. Boukhvalov}
\affiliation{Institute for Molecules and Materials, Radboud
University Nijmegen, NL-6525 ED Nijmegen, the Netherlands}
\affiliation{Institute of Metal Physics, Russian Academy of
Sciences, Ural Division, Ekaterinburg 620041, Russia}
\author{Yu. N. Gornostyrev}
\affiliation{Institute of Metal Physics, Russian Academy of
Sciences, Ural Division, Ekaterinburg 620041, Russia}
\affiliation{Institute of Quantum Materials Science, Ekaterinburg
620107, Russia}
\author{M. I. Katsnelson}
\affiliation{Institute for Molecules and Materials, Radboud
University Nijmegen, NL-6525 ED Nijmegen, the Netherlands}
\author{A. I. Lichtenstein}
\affiliation{Institut f\"ur Theoretische Physik, Universit\"at
Hamburg, 20355 Hamburg, Germany}

\date{\today}

\pacs{61.72.Ji, 75.20.En, 75.30.Et, 71.20.Be}

\begin{abstract}
Local perturbations of crystal and magnetic structure of
$\gamma$-iron near carbon interstitial impurity is investigated by
{\it ab initio} electronic structure calculations. It is shown
that the carbon impurity creates locally a region of ferromagnetic
(FM) ordering with substantial tetragonal distortions. Exchange
integrals and solution enthalpy are calculated, the latter being
in a very good agreement with experimental data. Effect of the
local distortions on the carbon-carbon interactions in
$\gamma$-iron is discussed.
\end{abstract}

\maketitle

Steel is a material playing unique role in our civilization. Its
main chemical composition is very simple, this is just iron and
carbon. Nevertheless, surprisingly, basic microscopic physics
determining phase and structural states of the steel is rather
poorly understood yet. In particular, mechanisms of development of
lattice instability and martensitic transformation at cooling down
of $\gamma$-Fe are still unknown. It is commonly accepted now that
magnetism is of crucial importance for the phase stability of iron
and its alloys~\cite{kaufman,hasegawa}, however, more or less
detailed theoretical studies have been carried out only for a
particular case of Invar Fe-Ni
alloys~\cite{wasserman,schilfgaarde,ruban}.

Commonly used steel is based on the low-temperature ($\alpha$) bcc
phase of iron. However, morphology of its microstructure which is
decisive for all practical applications is formed during the
quenching process from high-temperature fcc ($\gamma$) phase.
Kinetics of the $\gamma -\alpha$ transition is very sensitive to
carbon concentration. The state of carbon in $\alpha$-iron was a
subject of numerous
investigations~\cite{goldschmidt,jiang,domain}. The state of
carbon in $\gamma$-iron is much less understood, and even its
solution enthalpy calculated by state-of-art {\it ab initio}
approach strongly disagrees with the experimental data~\cite{lobo}
which is quite unusual.

The most probably, magnetic state of $\gamma$-iron is relevant for
structural properties of its alloys. Its magnetic state is
strongly frustrated which leads to existence of numerous
complicated magnetic structures with very close
energies~\cite{pinski,mryasov,antropov}. Role of the lability of
magnetic structure and the frustrations is discussed already in
the context of Invar problem~\cite{schilfgaarde,ruban}. Here we
present the results of {\it ab initio} calculations of the
electronic structure, lattice and magnetic properties of carbon
solid solution in  $\gamma$-iron. It turns out that the carbon
interstitial in octahedron void results in an essential local
magnetic polarization and strong lattice distortions which should
be taken into account, in particular, to obtain correct value of
the solution enthalpy.

We used the SIESTA package of first-principles electronic
structure calculations~\cite{siesta1,siesta2} with the generalized
gradient approximation for the density functional~\cite{perdew}.
Earlier the same approach has been successfully used to calculate
various properties of bulk and surface iron~\cite{siesta3} as well
as Fe clusters~\cite{siesta4}. To calculate exchange interactions
of the effective  Heisenberg model
\begin{equation}
H_{eff} = - \sum_{i,j} J_{i,j} {\bf S}_i {\bf S}_j \nonumber
\end{equation}
(${\bf S}_i$ are the classical spins defined by the direction and
magnitude of obtained magnetic moments) a standard density
functional approach has been used based on the ``magnetic force
theorem''~\cite{exchange}. We optimized first the structure and
then use the implementation of the Green's
function~\cite{GFLMTO,ldau} into the LMTO method~\cite{LMTO} to
calculate the effective exchange parameters which is not possible
in the framework of SIESTA. Justification of this approach will be
presented below, at the discussion of the Table \ref{tab2}.

\begin{table}
\caption{\\ Lattice parameters, magnetic moments, and energies of
different magnetic configurations per atom for $\gamma$-Fe with
atomic relaxation taken into account; numbers in parenthesis are
taken from Ref.\onlinecite{marcus}.} \label{tab1}
\begin{ruledtabular}
\begin{tabular}{cccc}
 & FM & AFM & AFMD  \\ \hline
$a$, {\AA} & 3.58 (3.45) & 3.44 (3.45) & 3.57 (3.49)\\
$c/a$ & 1.08 (1.18) & 1.09 (1.09) & 1.05 (1.09) \\
$M$, $\mu_B$ & 2.5 & 1.8 & 2.3 \\
$E-E_{AFMD}$, meV & 33 & 45 & 0 \\
\end{tabular}
\end{ruledtabular}
\end{table}

\begin{figure}[!htb]
\includegraphics[width=2.4 in]{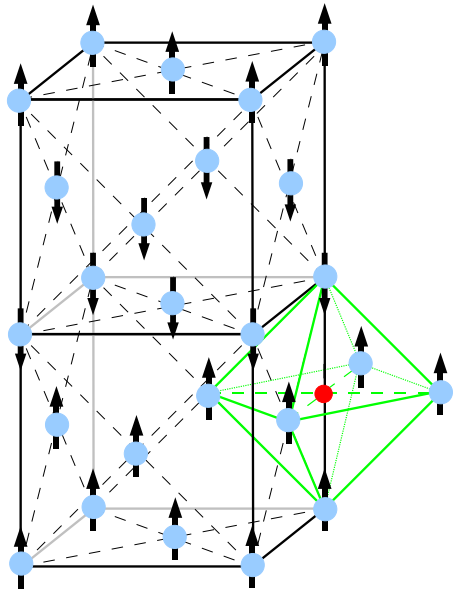}
\caption{\label{fig1}(color online) Fragment of crystal and
magnetic structure of $\gamma$-Fe for the AFMD magnetic ordering.
Carbon interstitial impurity in octahedral position is shown by
red (dark) circle.}
\end{figure}

To check accuracy of our approach we first studied structural
relaxation effects in pure $\gamma$-Fe which is known to result in
tetragonal deformations of initial fcc
lattice~\cite{marcus,marsman}. The computational results presented
in the Table \ref{tab1} demonstrate a reasonable agreement with
the previous calculations and with experimental data for thin
films of $\gamma$-Fe~\cite{marcus}. We have considered the
following magnetic structures: ferromagnetic (FM),
antiferromagnetic with the staggered magnetization in $<001>$
direction (AFM), and double antiferromagnetic (AFMD), or
``$++--$'' (see Fig. \ref{fig1}). The latter magnetic
configuration is one of most energetically favorable for
$\gamma$-Fe~\cite{antropov,schilfgaarde}.

\begin{figure}[!htb]
\includegraphics[width=3.3 in]{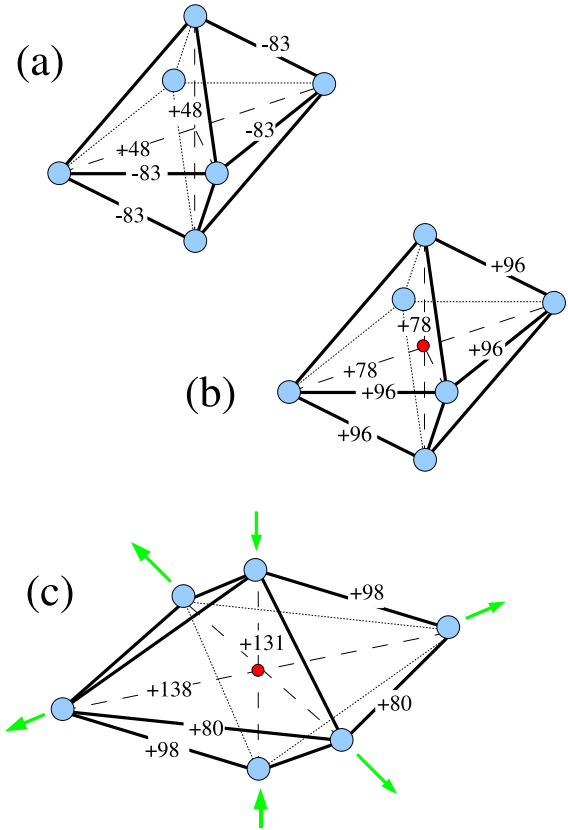}
\caption{\label{fig2}(color online) Exchange parameters (in K) for
different Fe-Fe pairs in original fcc lattice (a); in fcc lattice
with carbon interstitial impurity without (b) and with (c)
relaxation taken into account. Arrows indicate direction of atomic
displacements during the relaxation.}
\end{figure}

Further, we have performed calculations for Fe$_{32}$C supercell
with carbon in the octahedral void (Fig. \ref{fig1}). This
concentration is close to the eutectic point (3.6 at.\%) at the
phase diagram Fe-C~\cite{haasen} which is the most interesting
from the point of view of metallurgy. Calculations of the total
energy and magnetic moments, as well as optimization of positions
of all atoms in the supercell have been carried out for FM, AFM,
and AFMD structures (see Table \ref{tab2}). Calculated exchange
interactions for the first and second Fe--Fe bonds have shown in
Fig. \ref{fig2}a. These parameters agree well with previous
calculations for fcc iron~\cite{ruban}. Carbon in the octahedral
void, even without relaxation, changes the sign of
nearest-neighbor exchange parameters from AFM ($J_1$=-83 K) to FM
($J_1$=+96 K). Another effect is an essential increase of the
next-nearest-neighbor exchange parameter ($J_2$=48 K in Fig.
\ref{fig2}a and $J_2$=78 K in Fig. \ref{fig2}b). Interestingly,
the relaxation makes the next-nearest-neighbor interactions even
stronger than the nearest-neighbor ones (Fig. \ref{fig2}c).
Probably, the gain of magnetic energy related to this effect is
one of the driving mechanism of the local tetragonal distortion.
The main magnetic characteristics calculated in the SIESTA and in
the LMTO are similar which confirms that our exchange parameters
are reliable enough, at least, for qualitative discussions.

It turned out that, in contrast with the case of pure
$\gamma$-iron, the FM ordering has the lowest energy in the
presence of carbon. The exchange parameters~\cite{exchange}
calculated for the FM configuration presented in Fig. \ref{fig2}
also confirm that this magnetic configuration is stable. The
accuracy of the Heisenberg model estimated from the difference of
exchange parameters and values of magnetic moments in the FM and
AFMD state is in the limit of 25\% (see Table \ref{tab2}).

\begin{table}
\caption{\\ Lattice parameters,  tetragonal deformations, magnetic
moments for nearest neighbors ({\it nn}) and next nearest
neighbors ({\it nnn}), and total energy differences per iron atom
for Fe$_{32}$C unit cell; numbers in parenthesis calculated within
LMTO values}. \label{tab2}
\begin{ruledtabular}
\begin{tabular}{cccc}
 & FM & AFM & AFMD  \\ \hline
$a$, {\AA} & 3.73 & 3.56 & 3.61 \\
$c/a$, {\it nn} & 0.94 & 0.98 & 0.93 \\
$c/a$, {\it nnn}  & 0.99 & 1.04 & 1.00 \\
$c/a$, bulk & 0.97 & 1.04 & 1.00\\ \hline
$M$, $\mu_B$ {\it nn}  & 2.3 (2.1) & 1.6 (1.8) & 1.9 (1.9)\\
$M$, $\mu_B$ {\it nnn} &  2.8 (2.7) & 2.2 (2.4) & 2.6 (2.5)\\
$M$, $\mu_B$ bulk  &  2.7 (2.5) & 1.9 (2.2) & 2.1 (2.3)\\
$E-E_{FM}$, meV & 0 & 47 (29) & 16 (13) \\
\end{tabular}
\end{ruledtabular}
\end{table}

\begin{figure}[!htb]
\includegraphics[width=3.2 in]{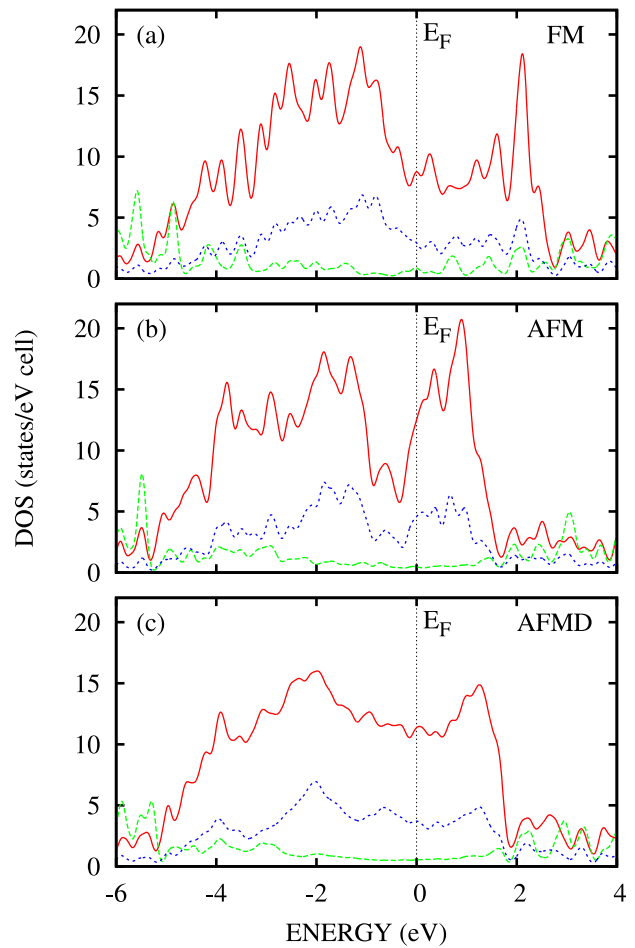}
\caption{\label{fig3} (color online) Total DOS of iron atoms
(solid red line), DOS of iron atoms from the first shell (dotted
blue line), and for carbon atom (dashed green line) for relaxed
Fe$_{32}$C supercell with carbon in octahedral void for different
magnetic configurations. }
\end{figure}

The mechanism of FM state stabilization by the carbon impurity can
be understood by investigation of Fe-C chemical bonding. We
present the density of states (DOS) for Fe in Fe$_{32}$C
supercell, together with the local DOS for carbon impurity, in
Fig. \ref{fig3}. One can see that for a broad energy interval
($\pm 2$eV) near the Fermi level the hybridization of the
$sp$-states of carbon with the $d$-states of iron is much more
pronounced for the FM state (Fig. \ref{fig3}a) than for both AFM
ones (Fig. \ref{fig3}b,c). This can lead to the energy
stabilization of FM states in fcc Fe-C alloys, which results in
positive exchange interactions even without structure relaxation
(Fig. \ref{fig2}b). The effect of anisotropic structural
relaxation increases formation of strong FM bonds (Fig.
\ref{fig2}c) and reduce the total iron DOS at the Fermi level
(Fig. \ref{fig3}a) in comparison with the original AFM states.

The solution enthalpy of carbon in $\gamma$-iron has been
calculated from the total energies of FM Fe$_{32}$C and of AFMD
fcc Fe (which have the lowest energies among trial magnetic
configurations), together with the ground state energy of
graphite. The result is 0.55 eV whereas experimental value is
about 0.4 eV~\cite{lobo}. Keeping in mind that {\it ab initio}
calculations without taking into account local distortions and
correct magnetic ground state give just a wrong sign for this
quantity~\cite{jiang} one can say that the agreement is rather
good. Actually, this is even better since our calculations have
been done for high enough carbon concentration and thus a
fictitious carbon-carbon interaction presents. Estimations of this
effect according to the standard elasticity
theory~\cite{khachaturyan} gives a value of order of 0.1 eV which
should be subtracted from our result.

We have done also calculations for the supercell Fe$_{108}$C with
the AFMD magnetic configuration. Starting from third coordination
sphere of the carbon atom the tetragonal deformation $c/a-1$ is
approximately 7\% which is close to the value for pure iron (see
Table \ref{tab1}). The corresponding values of local tetragonal
distortions for the first shell near the carbon impurity is 3\%
and for second shell is already 6\%. This means that respective to
the tetragonal deformation of the host carbon produces local
distortion of its nearest surrounding at approximately -4\% which
is a bit smaller than for Fe$_{32}$C supercell. The values of
magnetic moments (in $\mu_B$ units) for the first three
coordination spheres are 1.9, 2.5, and 2.6, respectively. Thus,
local tetragonal distortion around carbon atom can be considered,
indeed, as an effect of single impurity.

Interactions between carbon atoms via lattice distortions
(deformation interactions) determine decomposition and carbon
ordering processes in steel which are important for microstructure
formation~\cite{haasen}. Octahedral voids in bcc lattice are
asymmetric themselves which results in local tetragonal
distortions around carbon interstitial impurity and rather strong
deformation interactions. It is commonly accepted that the
deformation interactions in fcc host are much weaker since the
voids are symmetric and the interstitial carbons are considered as
purely dilatations centers~\cite{khachaturyan}. We demonstrate
that proper including the magnetic effects leads to the local
tetragonal distortions around carbon in $\gamma$-Fe of the same
order of magnitude as in $\alpha$-Fe and, thus, traditional views
on the importance of the deformation interactions in different
phases should be reconsidered.

These local deformations are intimately connected with the effect
of carbon on local magnetic configurations of iron. It turns out
that carbon changes signs of some exchange integrals from AFM to
FM. Similar effects of strong distortion dependence of exchange
interactions has been discussed earlier for fcc Fe-Ni
alloys~\cite{ruban}. We show here that in addition to above
mentioned distance dependence of effective exchange interaction
the effect of Fe-C chemical bonding is also important.

It should be noted, that our calculations are done for the
ground-state case whereas an interesting temperature interval for
Fe-C steels is above 10$^3$ K.  Nevertheless, the investigated
magnetic effects can be very important for understanding of
structural distortions in $\gamma$-iron alloy. Indeed, local
magnetic configurations and thus local distortions can survive
till relatively high temperatures. For the classical Heisenberg
model on fcc lattice, the mean-field estimation for the Curie
temperature is equal to the energy difference between FM and AFM
configurations which gives a value of order of 500 K (see Table
\ref{tab2}). Quantum effects for magnetic moments of order of 2
$\mu_B$ increases this estimation by a factor of
2~\cite{exchange,kotliar} which allows to assume that, at least,
up to 10$^3$ K local magnetic correlations will survive. This
temperature can be higher for higher concentration of carbon.
There are some direct experimental evidences that magnetic effects
are important for the austenite-to-ferrite ($\alpha - \gamma$)
transformation in steel~\cite{zhang}.

In conclusion, the complex magnetic state with strong tetragonal
distortions is predicted for $\gamma$-iron near carbon impurities.
The calculated exchange interactions show the strong tendency to
formation of local FM clusters. This effect changes drastically
carbon-carbon deformation interactions in $\gamma$-phase and thus
should be relevant for the martensitic transformations in steel.

\section*{Acknowledgements}

The work was supported by NWO (project 047.016.005) and by FOM
(Netherlands). D. W. B. acknowledges support from Funding by the
Research Council of the President of the Russian Federation (Grant
NSH-4640.2006.2). Y. N. G. acknowledges support by Russian Basic
Research Foundation (Grant 06-02-16557).


\begin{thebibliography}{99}
\bibitem{kaufman} L. Kaufman, E. V. Clougherty, and R. J. Weiss,
Acta Metal. {\bf 11}, 323 (1963).

\bibitem{hasegawa} H. Hasegawa and D. G. Pettifor, Phys. Rev.
Lett. {\bf 50}, 130 (1983).

\bibitem{wasserman} E. F. Wasserman, in: {\it Ferromagnetic
Materials}, ed. by K. H. J. Buschow and E. P. Wolfarth (North
Holland, Amsterdam, 1990), vol. 5, p. 237.

\bibitem{schilfgaarde} M. van Schilfgaarde, I. A. Abrikosov, and
B. Johansson, Nature {\bf 400}, 46 (1999).

\bibitem{ruban} A. V. Ruban {\it et al}, Phys. Rev. B {\bf 71}, 054402 (2005).

\bibitem{goldschmidt} H. J. Goldschmidt, {\it Interstitial Alloys} (Plenum Press, New York, 1967).

\bibitem{jiang} D. E. Jiang and E. A. Carter, Phys.
Rev. B {\bf 67}, 214103 (2003).

\bibitem{domain} C. Domain, C. S. Becquart, and J. Foct, Phys.
Rev. B {\bf 69}, 144112 (2004).

\bibitem{lobo} J. A. Lobo and G. H. Geiger, Metal. Trans. A {\bf 7}, 1359
(1976).

\bibitem{pinski} F. J. Pinski {\it et al}, Phys. Rev. Lett. {\bf 56}, 2096 (1986).

\bibitem{mryasov} O. N. Mryasov {\it et al},  J.
Phys.: Condens. Matter {\bf 3}, 7683 (1991).

\bibitem{antropov} V. P. Antropov {\it et al}, Phys. Rev. Lett. {\bf 75}, 729 (1995).

\bibitem{siesta1} E. Artacho {\it et al}, SIESTA, Version 1.3, 2004.

\bibitem{siesta2} J. M. Soler {\it et al}, J. Phys.:
Condens. Matter {\bf 14}, 2745 (2002).

\bibitem{perdew} J. P. Perdew, K. Burke, and M. Ernzerhof, Phys. Rev. Lett. {\bf 77}, 3865 (1996).

\bibitem{siesta3} J. Izquierdo {\it et al}, Phys. Rev. B {\bf 61}, 13639 (2000).

\bibitem{siesta4} A. V. Postnikov, P. Entel, and J. M. Soler, Eur.
Phys. J. D {\bf 25}, 261 (2003).

\bibitem{exchange} A. I. Liechtenstein, M. I. Katsnelson, and V. A.
Gubanov, J. Phys. F {\bf 14}, L125 (1984); Solid State Commun.
{\bf 54}, 327 (1985); A. I. Liechtenstein {\it et al}, J. Magn.
Magn. Mater. {\bf 67}, 65 (1987).

\bibitem{GFLMTO}  O. Gunnarsson, O. Jepsen, and O. K. Andersen,
Phys. Rev. B {\bf 27}, 7144 (1983).

\bibitem{ldau} V. I. Anisimov, F. Aryasetiawan, and A. I. Lichtenstein,
J.Phys.: Condens. Matter {\bf 9}, 767 (1997).

\bibitem{LMTO}
O. K. Andersen, Phys. Rev. B {\bf 12}, 3060 (1975).

\bibitem{marcus} P. M. Marcus, V. L. Moruzzi, and S. L. Qiu, Phys.
Rev. B {\bf 60}, 369 (1999).

\bibitem{marsman} M. Marsman and J. Hafner, Phys. Rev. B {\bf 66}, 224409
(2002).

\bibitem{haasen} P. Haasen, {\it Physical Matallurgy} (Cambridge
Univ. Press, Cambridge, 1986).

\bibitem{khachaturyan} A. G. Khachaturyan, {\it Theory of Structural Transformation in
Solids} (Wiley, New York, 1983).

\bibitem{kotliar} A. I. Lichtenstein, M. I. Katsnelson, and G.
Kotliar, Phys. Rev. Lett. {\bf 87}, 067205 (2001).

\bibitem{zhang} Y. Zhang {\it et al}, J. Magn. Magn. Mater. {\bf 294}, 267 (2005).
\end{thebibliography}
\end{document}